\documentclass[
    reprint,
    preprintnumbers,
    superscriptaddress,
    nofootinbib,
     amsmath,amssymb,
     aps,
     prd,
    floatfix,
    ]{revtex4-2}
    
    \usepackage{graphicx}
    \usepackage[caption=false]{subfig}
    \usepackage[usenames,dvipsnames]{xcolor} 
    \usepackage[utf8]{inputenc}
    \usepackage{color}
    \usepackage{hyperref}
    \usepackage[normalem]{ulem} 
    \usepackage{physics}

    \usepackage{enumitem}
    \usepackage{comment}
    \usepackage{bm}
    \hypersetup{
      colorlinks   = true, 
      urlcolor     = blue, 
      linkcolor    = blue, 
      citecolor   = red 
    }
    \allowdisplaybreaks[1]
    \graphicspath{{figures/}}

\usepackage{float}

\newcommand{\mpl}{M_\mathrm{Pl}}

\begin{document}

\title{Preheating and oscillon formation in Einstein-scalar-Gauss-Bonnet gravity}

\newcommand{\QMUL}{Centre for Geometry, Analysis and Gravitation, School of Mathematical Sciences, Queen Mary University of London,
Mile End Road, London E1 4NS, United Kingdom}
\newcommand{\MIT}{Center for Theoretical Physics -- A Leinweber Institute, Massachusetts Institute of Technology, Cambridge, MA 02139, USA}

\author{Areef Waeming}
\email{a.waeming@qmul.ac.uk}
\affiliation{\QMUL}

\author{Josu C. Aurrekoetxea}
\email{jaurreko@mit.edu}
\affiliation{\MIT}

\author{Katy Clough}
\email{k.clough@qmul.ac.uk}
\affiliation{\QMUL}

\author{Pau Figueras}
\email{p.figueras@qmul.ac.uk}
\affiliation{\QMUL}

\author{\'Aron D. Kov\'acs}
\email{a.kovacs@qmul.ac.uk}
\affiliation{\QMUL}


\begin{abstract}
Non-perturbative processes in the early universe may create overdense structures in scalar fields like the inflaton, called oscillons. In this work, we explore whether the leading order higher derivative contributions to the scalar-tensor theory change the formation and growth of these structures, and investigate the limits in which the effective field theory (EFT) description breaks down. We find that whilst the properties of the oscillons are not significantly modified, and black holes do not generically form, for large couplings the period of formation can result in the evolution leaving the regime of validity of the EFT, at which point predictivity is lost and the next order terms in the EFT should become relevant. If the oscillons survive their formation, they tend to be stable and the EFT corrections remain bounded. The EFT breakdown is triggered by large curvature terms in the metric in the densest regions of the oscillon, meaning that approximations of such modified theories that neglect the local backreaction and non-linear dynamics of the fields may miss important effects.
\end{abstract}

\maketitle

\section{\label{sec:intro}Introduction}
Cosmic inflation \cite{Guth:1980zm, Linde:1981mu, Albrecht:1982wi, Starobinsky:1980te,Linde:1983gd} is an early phase of accelerated expansion that resolves several fundamental cosmological issues, such as the horizon and flatness problems, as well as sourcing a spectrum of scale-invariant perturbations consistent with those observed in the Cosmic Microwave Background (CMB) on large scales \cite{2011ApJS..192...18K,Planck:2018vyg}. Although single-field slow-roll inflation is consistent with current observational data \cite{Planck:2018jri}, recent results from the Atacama Cosmology Telescope and the South Pole Telescope \cite{AtacamaCosmologyTelescope:2025blo,AtacamaCosmologyTelescope:2025nti,SPT-3G:2025bzu} have led to tensions with the favoured $\alpha$-attractor models \cite{Kallosh:2013yoa,Kallosh:2013maa,Kallosh:2014rga}, of which Starobinsky inflation \cite{Starobinsky:1980te} is one example. Many proposals have been considered to reconcile these tensions, including accounting for the effect of the leading order higher derivative terms in the action \cite{Oikonomou:2021kql,Chakraborty:2018scm,Yogesh:2024zwi,Yogesh:2025wak,Zhu:2025twm,Odintsov:2025wai,Zahoor:2025nuq,Addazi:2025qra,Odintsov:2025bmp}. The Starobinsky model itself was originally motivated as an effective theory arising from a quadratic Ricci scalar term ($\propto R^2$) in the Einstein-Hilbert action. Here we still consider the potential term to represent the effect of UV physics that has been integrated out, and add the leading order derivative interactions in addition to it. The addition of these terms is motivated by the idea that classical General Relativity (GR) is the low-energy limit of some other UV-complete theory, and therefore there should be additional higher-derivative terms in the action that become relevant above some (as yet unknown) energy scale. The leading order corrections result in a non-trivial coupling between the Gauss-Bonnet invariant and the scalar field  -- they are Einstein-scalar-Gauss-Bonnet (EsGB) models \cite{	Lovelock:1971yv,Zwiebach:1985uq,Boulware:1985wk}, members of the Horndeski class \cite {Horndeski:1974wa} -- which could affect the dynamics of the early universe. 

If inflaton dynamics are modified, the cut-off scale would be close to the order of the inflationary scale $\ell\sim 1/H_\mathrm{inf}$, which can take a range of values, but for typical high-scale inflation $1/H_\mathrm{inf} \sim 10^5 \ell_\mathrm{Pl}$. Such inflationary-scale modifications should also play a role during the subsequent reheating and/or preheating phases \cite{Kofman:1994rk,Kofman:1997yn,Albrecht:1982mp, Traschen:1990sw} (see \cite{Amin:2014eta} for a review), in which the accelerated inflationary period terminates and the energy stored in the inflaton field is transferred to Standard Model particles. The physics of this period is not well-constrained by observational data, and so the detailed dynamics remain uncertain and may involve both perturbative and non-perturbative decay channels of the inflaton field. In preheating scenarios, which occur generically for $\alpha$-attractor models, the inflaton field undergoes coherent oscillations that resonantly amplify specific field modes around the mass scale of the potential. This process can lead to the formation of localised, long-lived scalar configurations known as oscillons \cite{Bogolyubsky:1976nx,Bogolyubsky:1976sc,Gleiser:1993pt,Copeland:1995fq,Kasuya:2002zs,Saffin:2006yk,Hertzberg:2010yz,Salmi:2012ta,Gleiser:2019rvw,Antusch:2019qrr,Ibe:2019vyo,Zhang:2020bec,vanDissel:2023zva} (see \cite{Zhou:2024mea} for a recent review). Such objects are of cosmological interest, as they may act as seeds for primordial black hole formation and/or generate characteristic signatures in the stochastic gravitational wave background. Preheating has been studied using analytical and numerical techniques \cite{Felder:2000hq,Felder:2006cc,Frolov:2008hy,Amin:2010jq, Amin:2010dc, Amin:2011hj, Amin:2013ika, Lozanov:2014zfa, Antusch:2015ziz, Antusch:2015vna, DeCross:2015uza,DeCross:2016fdz,DeCross:2016cbs,Kim:2017duj,Lozanov:2017hjm, Musoke:2019ima, Niemeyer:2019gab,Nguyen:2019kbm,Martin:2019nuw,Figueroa:2020rrl,Figueroa:2021yhd,Eggemeier:2020zeg,Iarygina:2020dwe,Sang:2020kpd,Martin:2020fgl,vandeVis:2020qcp,Kost:2021rbi,Garcia:2021iag,Eggemeier:2021smj,Kim:2021ipz,Figueroa:2022iho,Joana:2022uwc,Mahbub:2023faw,Piani:2023aof,Shafi:2024jig,Sui:2024grm,Jia:2024fmo,Piani:2025dpy,Baeza-Ballesteros:2025tme,Li:2025ioq,del-Corral:2025fzz,Martinez:2025ana,Gu:2026ajw,Lozanov:2026dgu,Laverda:2026slq}, but in most cases these works only consider perturbations on a homogeneous Friedmann–Lemaître–Robertson–Walker (FLRW) background, or modifications to the homogeneous background itself, particularly once higher derivative corrections are included, e.g., \cite{Oikonomou:2024jqv,Zahoor:2025nuq,Odintsov:2025bmp}. In this work, we consider whether such approximations remain valid in a fully non-perturbative evolution with higher derivative corrections. In particular, we are interested in whether the properties of the oscillons are significantly modified, and whether the impact on the local spacetime curvature of the overdensities becomes relevant and triggers a breakdown of the EFT.

Several works have studied preheating with gravitational backreaction. In \cite{Lozanov:2019ylm, Amin:2019ums} the effect of oscillons was treated with a Newtonian potential with a Poisson equation sourced by the energy density. Further works \cite{Easther:1999ws,Finelli:2000gi,Bastero-Gil:2007lsx,Bastero-Gil:2010tpb,Bastero-Gil:2011jyw, Kou:2019bbc, Giblin:2019nuv, Kou:2021bij,Aurrekoetxea:2023jwd,Adshead:2023mvt} have also included the effects of the gravitational backreaction at the fully non-linear level of general relativity using numerical relativity (NR).
The density of oscillons that results is model dependent, but it was shown in Ref. \cite{Aurrekoetxea:2023jwd} that for the simplest single field models consistent with observational constraints their compactness is generally too low to trigger black hole formation. Multi-field models offer more favourable conditions, and recent work has suggested that black holes may form in kinetic preheating models \cite{Adshead:2025gka}. See Ref. \cite{Aurrekoetxea:2024ypv} for a review of the application of NR to study cosmology.

Studying cases of higher derivative theories with numerical relativity simulations is at a very early stage, since until recently it was not clear how to render the equations in a form that admits a well-posed initial value problem. Simulations to constrain EFTs of gravity with corrections that turn on at higher curvature scales have focussed on compact objects \cite{Doneva:2024ntw,Doneva:2023oww,Thaalba:2024crk,Thaalba:2024htc,Thaalba:2023fmq,Franchini:2022ukz,East:2020hgw,East:2021bqk,Corman:2024vlk, Corman:2022xqg,Corman:2024cdr,Ripley:2019aqj,Ripley:2019aqj,AresteSalo:2022hua,AresteSalo:2023mmd,R:2022hlf}, with the exception of a recent work looking at their effects on inflation \cite{Brady:2025zxp}. The best constraints on EsGB models are currently set to the order of kilometre length scales by observations of the black hole-neutron star gravitational wave (GW) signal GW230529 using Post-Newtonian results for EsGB \cite{Sanger:2024axs}, with shorter length scales unconstrained. To probe higher curvatures than this with compact objects would require less massive black holes (which would need to be primordial in origin). The consistency of higher derivative effects in the early universe therefore has the potential to set more stringent constraints, provided we understand the dynamics. However, as pointed out in Ref. \cite{deRham:2018red}, we note that constraints should not be extrapolated naively across regimes since these bounds depend on the characteristic frequencies and background field configurations of the observations, and may be altered near the EFT cut-off.

Throughout the paper, we are using the following conventions. The indices indicated by the Greek letter $\mu,\,\nu\,,...$ represent 4-dimensional spacetime, which runs from 0 to 3. The Roman indices $i,\,j\,,...$ denote the spatial part of spacetime runs from 1 to 3. The derivative with respect to time is indicated by overdot, $(~\dot{}~)$. In the code we use geometric units where $8\pi G=c=1$, but in the paper we restore factors of the reduced Planck mass $M_\mathrm{Pl}$ when quoting values.

\section{Methodology}

\subsection*{Scalar-Tensor theory}

General relativity can be viewed as a low energy EFT, valid below a certain cut-off scale, in which higher-derivative operators encode the effects of unknown ultraviolet physics at low energies. 

In the EFT of single field inflation the leading corrections to the Lagrangian of the minimally coupled Einstein-scalar-field theory are operators with four derivatives. The most general set of such operators is detailed in \cite{Weinberg:2008hq}, but most of these operators are redundant in the sense that they are either total derivatives or they can be eliminated by using field redefinitions. The effective action simplifies even more if we additionally assume that our theory is invariant under spacetime parity transformations. The effective action is then
\begin{eqnarray}\label{eq:action}
S = \int d^4x \sqrt{-g}\left[ \frac{M_\mathrm{Pl}^2}{2}R -V(\phi) + X + g_2(\phi)X^2 \right. \nonumber \\  
+ \left. ~ f(\phi)\mathcal{R}^{\mathrm{GB}}\right] ~,
\end{eqnarray}
where 
\begin{equation}
    X = - \frac{1}{2}(\nabla_\mu\phi)(\nabla^\mu \phi)~,
\end{equation}
\begin{equation}
    \mathcal{R}^{\mathrm{GB}} = R^2-4R_{\mu\nu}R^{\mu\nu}+R_{\mu\nu\rho\sigma}R^{\mu\nu\rho\sigma} ~,
\end{equation}
and $g_2(\phi)$ and $f(\phi)$ are smooth functions of the scalar field $\phi$. In the case considered in this work, we set $g_2(\phi)=0$ and we consider only the Einstein-scalar-Gauss-Bonnet contribution.
Here $M_\mathrm{Pl} = \sqrt{1/8 \pi G}$ is the reduced Planck mass.

The equation of motion for the scalar field is then a modified version of the Klein-Gordon equation
\begin{eqnarray}
    \nabla_{\mu }\nabla^{\mu }\phi +\mathcal{R}^{\mathrm{GB}} f'(\phi) -  V'(\phi) = 0 ~, \label{eq:modKG}
\end{eqnarray}
where we see that curvature of the spacetime can now act as a source of the scalar field.

\begin{figure}[t]
    \includegraphics[width =\linewidth]{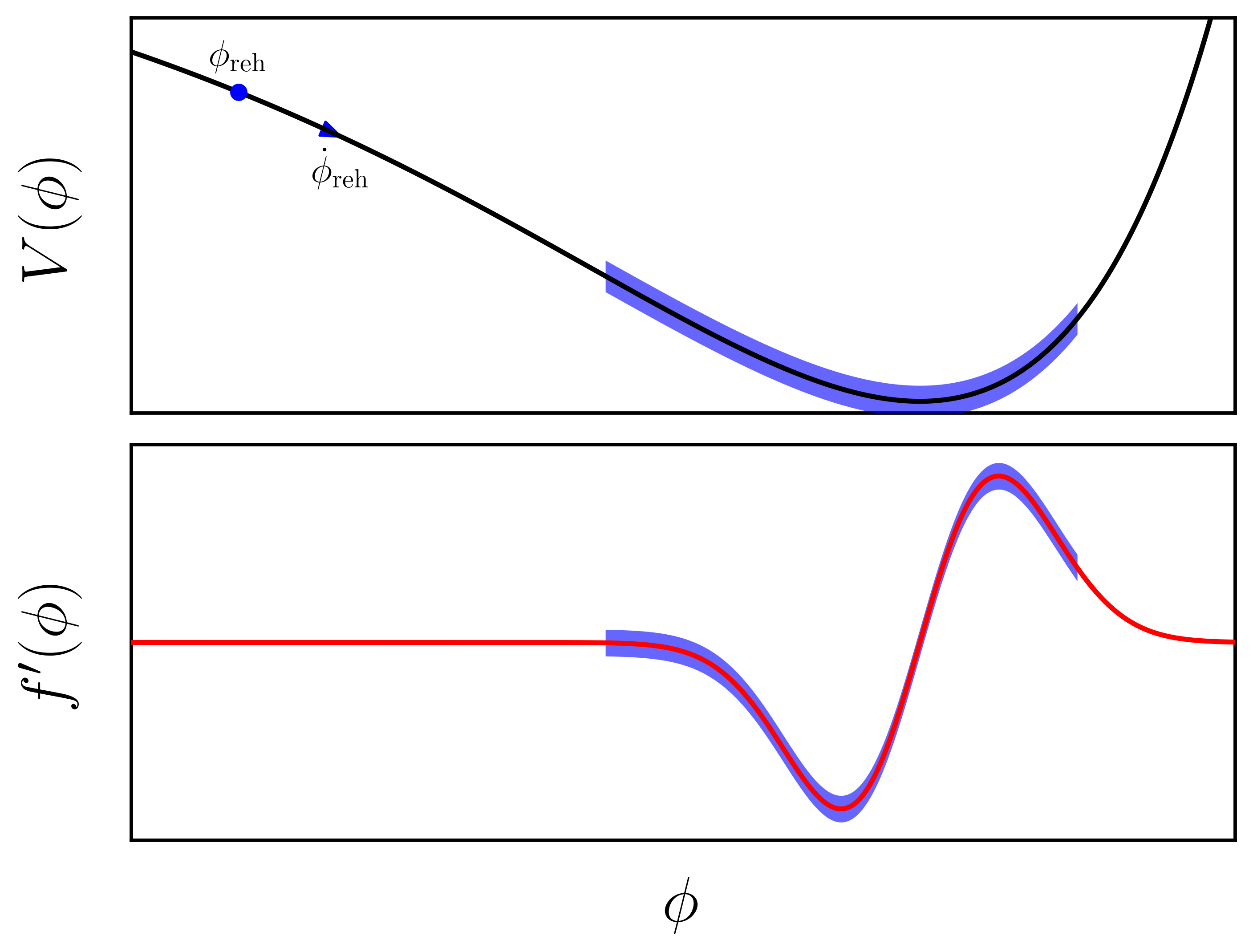}
    \caption{Scalar potential $V(\phi)$ and the derivative of coupling function $f'(\phi)$ which determines the strength of the modifications to GR. We start our simulations at the end of inflation at the point $\phi_{reh}$, with a velocity towards the minimum of the potential $\dot\phi_{reh}$. The modifications are chosen to be localised around the reheating part of the potential explored by the oscillon (highlighted in blue), and to be suppressed at $\phi_{reh}$.}
    \label{fig:pot_shape}
\end{figure}

\subsection*{Scalar field potential and coupling function}

In this work we follow closely the set up of \cite{Aurrekoetxea:2023jwd}, which studied $\alpha$-attractor models of the form\footnote{The $\alpha$-attractor potentials have an asymmetric shape with a plateau on one side and a steep wall on the other. The deviation from a quadratic form near the minimum provides the attractive self-interactions of the scalar field that enhances oscillon formation.}
\begin{eqnarray}
    V(\phi) = \frac{m^2 \mu^2}{2} \left( 1- e^{\phi/\mu}\right)^2 ~. \label{eq:pot}
\end{eqnarray}
Here $\mu$ parametrises whether the model is small or large field (how large the inflationary part of the plateau is) and $m$ sets the energy scale of the oscillations during preheating. The form of the potential is illustrated in Fig.~\ref{fig:pot_shape}. In this work we focus on the specific case with $\mu = 0.06 M_\mathrm{Pl}$, which was an intermediate model for the work of \cite{Aurrekoetxea:2023jwd}. In our simulations we set $m=1$ but this remains a free scale in the problem and so our results will be quoted in terms of $m$ where appropriate.

One can show that the modes of small perturbations above the homogeneous background in Fourier space evolve according to
\begin{eqnarray}
    \delta \ddot{\phi}_k +\left(k^2 + f''(\phi(t)) \mathcal{R}^{\mathrm{GB}} + V''(\phi(t))\right)\delta \phi_k(t)=0 ~. \label{eq:fieldGB_eom}
\end{eqnarray}
Therefore the Gauss-Bonnet coupling $f(\phi)$ will modify the modes that are excited, and their amplification and time dependence, but in a way that depends on the local value of the Gauss-Bonnet curvature invariant.

One of the simplest choices for $f(\phi)$ is a shift-symmetric coupling $f(\phi) = \lambda \phi$, where $\lambda$ is a coupling constant of the $\mathcal{R}^{\mathrm{GB}}$ term. However, the second derivative of this term vanishes, and therefore should not have a strong effect in the formation of oscillons c.f. Eq. \eqref{eq:fieldGB_eom}. In this work we study a coupling that is quadratic to leading order, specifically
\begin{eqnarray}
f(\phi) = \frac{\lambda}{2 \beta} \left[1-e^{- \beta \phi^2}\right], \label{exp_coupling}
\end{eqnarray}
where $\lambda$ has units of length$^2$ and sets the strength of the EsGB correction and $\beta=5000 M_\mathrm{Pl}^{-2}> 1/\mu^2$ is chosen to suppress the correction at the scale at the end of inflation (the point at which our simulation of preheating begins, $|\phi| \sim \mu$). The goal is for the coupling to smoothly transition between a flat part where inflation happens and a quadratic part around the minimum of the potential. This choice therefore allows us to isolate the leading order effect of the modification on the preheating period, separate from the modification it may make to the end of inflation. Although this is not a unique choice, it is a reasonably well-motivated candidate to study the effects. We expect that other choices with the same limits will give similar qualitative results. 
An EsGB-like modification to GR during an approximately homogeneous inflationary period can be represented by a modified scalar potential, since the homogeneous corrections dominate, as shown in \cite{Brady:2025zxp}. This also has the computational advantage that the equations reduce to their GR counterparts for the initial data and therefore we can start in the same initial state as that of \cite{Aurrekoetxea:2023jwd}. The forms of the coupling function and the potential are illustrated in Fig.~\ref{fig:pot_shape} -- note that they cannot be trivially added since the former comes with a factor of the local curvature (which varies in space and time) in the equation of motion. In this work we vary $\lambda$ to be positive and negative, and quote its strength in terms of the dimensionless curvature scale $\tilde{L} \equiv \sqrt{|\lambda|} H_\mathrm{reh}$, which compares the cut-off scale of the theory to the Hubble scale at the start of reheating.

\subsection*{Initial data}

The metric is decomposed following the Arnowitt–Deser–Misner (ADM) formalism 
\begin{eqnarray}
    ds^2 = -\alpha^2 dt^2 + \gamma_{ij}(dx^i + \beta^i dt)(dx^j + \beta^j dt)\,. \label{eq:ADMmetric}
\end{eqnarray}
Here $\alpha$, $\beta^i$, and $\gamma_{ij}$ are the lapse function, shift vector and 3-dimensional spatial metric, respectively. We also evolve the extrinsic curvature tensor $K_{ij}=\partial_{t} \gamma_{ij} + 2 D_{(i} \beta_{j)}$, which can be decomposed into a trace $K=\gamma^{ij}K_{ij}$  and a traceless part $A_{ij}$. 

For the initial data, we follow the same procedure as in Ref. \cite{Aurrekoetxea:2023jwd}. The initial data is chosen to be compatible with the Cosmic Microwave Background (CMB) observations from the Planck 2018 data \cite{Planck:2018jri}. We choose inflation to last 50 $e$-folds and the initial background values of the scalar field $\phi_\mathrm{reh} = \phi(t_\mathrm{reh})$ and its momentum $\dot{\phi}_\mathrm{reh} = \dot{\phi}(t_\mathrm{reh})$ for reheating are assigned corresponding to values at the end of inflation. As observed in CMB data, the scalar power spectrum,
\begin{eqnarray}
    \Delta^2_{\mathcal{R}}= \frac{H^2_\mathrm{inf}}{8 \pi^2 M^2_\mathrm{Pl} \epsilon(\phi_\mathrm{inf}) } \approx 2 \times 10^{-9}\label{eq:power_spec}
\end{eqnarray}
were used to generate the field fluctuations with fixed mass $m$. Note that in Eq. \eqref{eq:power_spec}, $\mathcal{R}$ refers to the curvature perturbation and the variables subscripted by ``inf'' are values at the beginning of inflation. As above, we focus on the specific case with $\mu = 0.06 M_\mathrm{Pl}$, which yields
\begin{eqnarray}
    \phi_\mathrm{inf} &=& -6.13988 \times 10^{-1} \,\,[M_\text{Pl}] \nonumber \\
    \phi_\mathrm{reh} &=& -8.64102 \times 10^{-2} \,\,[M_\text{Pl}], \nonumber\\
    \dot{\phi}_\mathrm{reh} &=& 3.23761 \times 10^{-2} \,\,[m M_\mathrm{Pl}], \nonumber \\
    H_\mathrm{reh} &=& 2.28934 \times 10^{-2} \,\,[m], \nonumber \\
    H^{-1}_\mathrm{reh} &=& 43.6897\,\,[m^{-1}]. \nonumber
\end{eqnarray}
We choose the length of our simulation domain to be $L=64 m^{-1} \gtrsim  H^{-1}_\mathrm{reh}$.
The scalar field contains perturbations
\begin{align}
    \phi(\textbf{x}) &= \phi_\mathrm{reh} +\delta\phi(\textbf{x}) \\ \dot{\phi}(\textbf{x})&=\dot{\phi}_\mathrm{reh}+\delta\dot{\phi}(\textbf{x}).
\end{align}
which, following the method presented in \cite{Figueroa:2020rrl}, assume a random Gaussian field, with sub-horizon scalar perturbations given by quantum vacuum fluctuations, i.e.
\begin{eqnarray}
    \mathcal{P}(k) &=& \frac{\xi}{2a^2_\mathrm{reh}\omega^2_k}, \\
     \langle \delta\phi_\textbf{k} \delta\phi_{\textbf{k}'} \rangle &=& (2 \pi)^3 \mathcal{P}(k) \delta(\textbf{k}-\textbf{k}'). \label{eq:scalar_pert_spec}
\end{eqnarray}
Here we define $a(t_\mathrm{reh})=1$ and $\omega_k = \sqrt{k^2+a^2_\mathrm{reh}V''(\phi_\mathrm{reh})}$. We introduce both ultraviolet and infrared cutoffs given by $k_\text{UV}=4 k_\text{IR} = 4(2 \pi/L)$. The variance of scalar perturbation is determined by 
\begin{eqnarray}
    \langle \delta\phi^2 \rangle = \int d\,\text{log}\,k \frac{k^3}{2 \pi^2} \mathcal{P}(k). \label{eq:var_field_pert}
\end{eqnarray}
We choose $\xi=10^4$ so that $\langle \delta\phi^2\rangle \approx 10^{-8} \mpl^2$. This represented an intermediate case in our previous work, giving oscillons of compactness $\mathcal{C} \sim 7.5\times 10^{-3}$. 

Due to the high computational cost of our simulations, we restricted our focus to a range of modes as discussed above, which should capture the relevant dynamics. Our study is set up to focus on the scales most amplified by parametric resonance in GR. Other wavelength modes are less amplified (at least in GR), so the modified gravity effects coming from the curvature terms should be smaller. Therefore, we do not expect that resolving longer or shorter length scales would change our main conclusion that modified gravity alters the physical properties of the oscillons very little up until the breakdown of the theory. Furthermore, while different random field realisations introduce slight variations in the exact positions, masses, and compactnesses of the oscillons, the qualitative behaviour remains robust \cite{Aurrekoetxea:2023jwd}. Unlike scenarios relying on rare, extreme fluctuations, our analysis focuses on the generic overdensities naturally produced by non-linear resonance in typical realisations.

The inhomogeneities in both the field and conjugate momentum source a non-trivial energy-momentum tensor for which the Hamiltonian and momentum constraints need to be solved. As in Ref. \cite{Aurrekoetxea:2023jwd}, we fix the initially flat three metric $\gamma_{ij}$ as $\delta_{ij}$, and solve the constraints for the trace of the extrinsic curvature tensor $K$ and the traceless part $A_{ij}$ using the CTTK method \cite{Aurrekoetxea:2022mpw} implemented in \texttt{GRTresna} \cite{Aurrekoetxea:2025kmm}. Note that as $f'(\phi_\mathrm{reh})\approx 0$, the initial data is the same for both GR and EsGB, so that the additional terms that arise from modified gravity do not need to be included in the initial data solution \cite{Brady:2023dgu} (although they will be excited dynamically during the subsequent evolution).

\subsection*{Evolution}

To obtain a well-posed initial value problem for \eqref{eq:action} and to solve the equations of motion of this theory numerically, it is necessary to find an appropriate gauge choice and gauge-fixing procedure \cite{Kovacs:2020pns,Kovacs:2020pns,AresteSalo:2022hua,AresteSalo:2023mmd}. Stable numerical simulations further require the addition of suitable constraint-damping terms, see \cite{AresteSalo:2022hua,AresteSalo:2023mmd} for more details. The gauge-fixed equations of motion we solve may be written as follows:
\begin{align}
    &R^{\mu\nu}-\textstyle\frac{1}{2}Rg^{\mu\nu}-\hat P_\alpha^{~\beta\mu\nu}\nabla_\beta C^\alpha \\
    &+ \kappa_1 [n ^{(\mu}C^{\nu)}+\textstyle\frac{1}{2}\kappa_2 n^\alpha C_\alpha g^{\mu\nu}] = T^{\phi~\mu\nu} - 4\mathcal{H}^{\mu\nu}, \nonumber\\
    &\square \phi[1+2g_2(\phi)X]-V'(\phi)-3X^2g_2'(\phi)\\
    &-2g_2(\phi)(\nabla^\mu\phi)(\nabla^\nu\phi)\nabla_\mu\nabla_\nu\phi=-f'(\phi)\mathcal{R}^{\mathrm{GB}}\nonumber
\end{align}
where
\begin{align}
    &T^\phi_{\mu\nu} = \textstyle\frac{1}{2}\{(\nabla_\mu\phi)(\nabla_\nu\phi)(1+2g_2(\phi)X)\\
    &+g_{\mu\nu}[g_2(\phi)X^2+X-V(\phi)]\},\nonumber\\
    &\hat P_\alpha^{~\beta\mu\nu}=\delta_\alpha^{~(\mu}\hat g^{\nu)\beta}-\textstyle\frac{1}{2}\delta_\alpha^{~\beta}\hat g^{\mu\nu}, \\
    &\mathcal{H}_{\mu\nu}=2R_{~(\mu}^{\rho}\mathcal{C}^{\mathstrut}_{\nu)\rho}-\mathcal{C}(R_{\mu\nu}-\textstyle\frac{1}{2}Rg_{\mu\nu})-\textstyle\frac{1}{2}R\mathcal{C}_{\mu\nu} \\
    & +  \mathcal{C}^{\alpha\beta}(R_{\mu\alpha\nu\beta}-g_{\mu\nu}R_{\alpha\beta}), \nonumber\\
    &\mathcal{C}_{\mu\nu}\equiv f'(\phi)\nabla_\mu\nabla_\nu\phi+ f''(\phi)(\nabla_\mu\phi)(\nabla_\nu\phi), \\
    &\mathcal{C}^\mu=H^\mu+\tilde g^{\rho\sigma}\Gamma^\mu_{\rho\sigma}.
\end{align}
$H^\mu$ are the source functions that parametrise the underlying coordinate freedom of the theory, and $\mathcal{C}\equiv g^{\mu\nu}\mathcal{C}_{\mu\nu}$. The auxiliary metrics are given by
\begin{align}
    \tilde g^{\mu\nu}=g^{\mu\nu}-a(x)n^\mu n^\nu,~~~
    \hat g^{\mu\nu}=g^{\mu\nu}-b(x)n^\mu n^\nu,
\end{align}
where $n^\mu$ is the unit normal to surfaces of constant time, $a(x)$ and $b(x)$ are chosen such that $0<a(x)<b(x)$ or $0<b(x)<a(x)$. In this work, we simply choose 0.2 and 0.4, respectively. The damping coefficients are set as $\kappa_1 = 1$, $\kappa_2 = -0.1$. 

To study inflationary preheating in EsGB, we evolve the system of equations using \texttt{GRFolres} \cite{AresteSalo:2023hcp}. This code uses the above formulation with a modified version of the puncture gauge \cite{AresteSalo:2023mmd}, such that the evolution of the lapse is given by
\begin{eqnarray}
    \partial_t \alpha = \beta^i\partial_i \alpha - \frac{2\alpha}{1+a(x)}(K-\overline{K}-2\Theta), \label{eq:mod_gauge}
\end{eqnarray}
Note that for this cosmological setup, we subtract the (proper volume averaged) mean value of the trace of the extrinsic curvature $\overline{K}$ so that the evolution in homogeneous regions is approximately FLRW background with the lapse remaining close to 1, as in \cite{Aurrekoetxea:2023jwd, Kou:2019bbc}. We also tried the approach suggested in \cite{Doherty:2025oui} in which the local density was used instead and found similar results, but it was slightly less stable during the collapse of the overdensities to form the oscillon and led to some simulations becoming unstable before leaving the EFT regime. We note that finding a good NR gauge that is well-adapted to cosmological simulations remains an open problem. In this work we experimented with several options and found similar qualitative results in all cases. Our diagnostics give us confidence that the breakdown we observe is a physical one, due to the breakdown in the EFT, rather than gauge issues.
That is to say, while gauge choice influences exactly when hyperbolicity is lost, the breakdown coincides with the EFT entering the strong coupling regime, confirming it as a genuine physical consequence of the modifications to GR rather than simply a poor choice of gauge. However, the exact critical coupling identified near $\tilde{L} \sim  0.180$ should be treated as an order-of-magnitude estimate that depends on specific model parameters.

\section{Diagnostics}\label{sec:diagnostics}

In order to quantify the differences with GR, we track the following diagnostic quantities as we vary the coupling strength $\tilde{L}$.

\begin{figure}[t]
    \includegraphics[width =\linewidth]{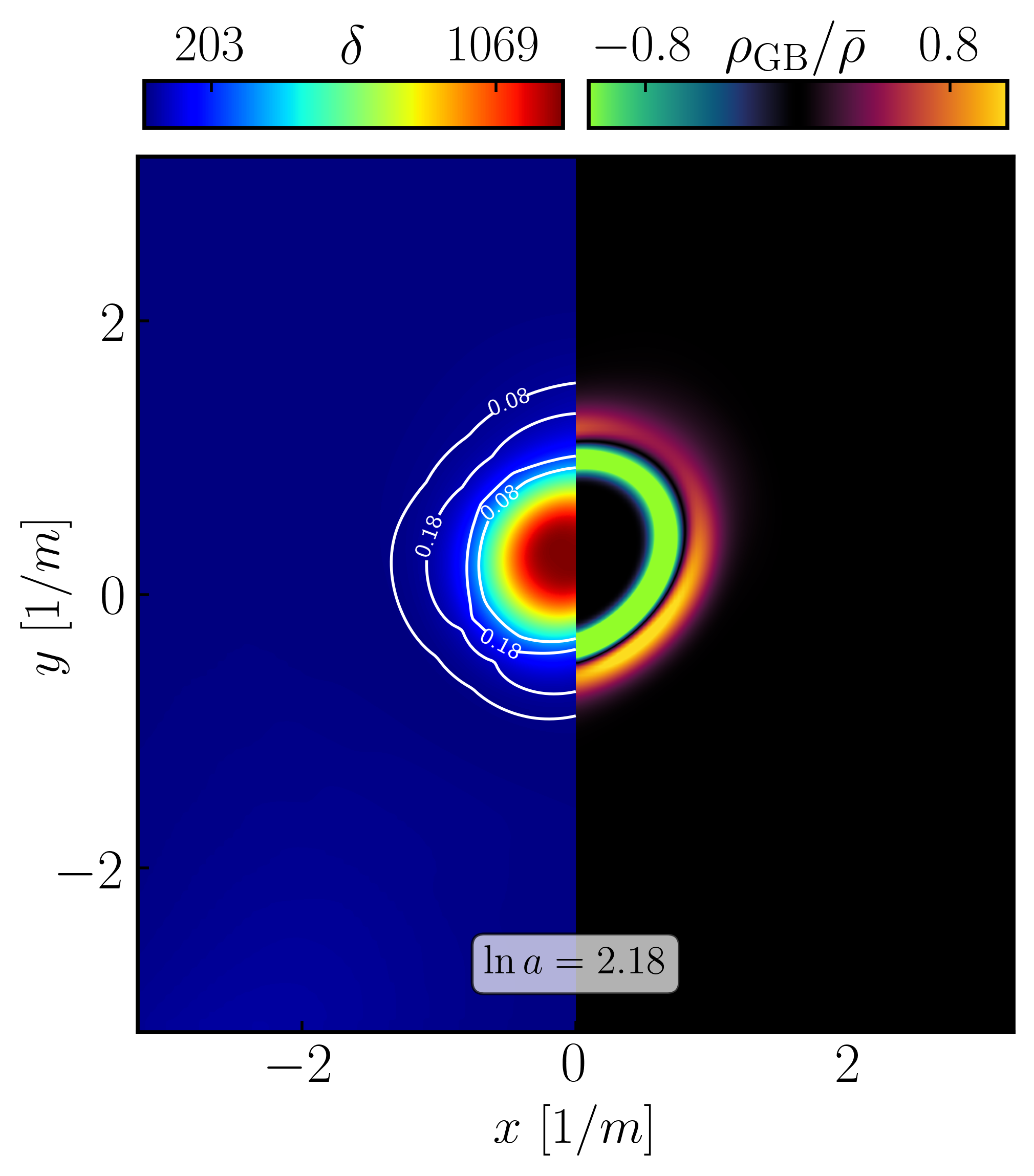}
    \caption{A snapshot of the density contrasts in the $x-y$ plane at a late time from the simulation with a positive $\lambda$ and $\tilde{L}=0.177$, illustrating the oscillon that has formed. The left panel depicts the total density contrast with the white lines being contours of the weak coupling condition ($\sqrt{|f'(\phi)|}/L$ with $L$ given by Eq. \eqref{eq:L_max}), and the right the Gauss-Bonnet contribution as defined in the main text. We see that the changes in the effective energy density due to the EsGB terms are localised around the walls of the oscillon, being sourced by the gradients of the scalar field, with both positive and negative contributions. In the case of the opposite sign of $\lambda$, the positive regions become negative and vice-versa.}
    \label{fig:snapshot_rho_phi_v_GB}
\end{figure}

\subsection*{Density contrast}
The \textit{density contrast} measures the over or underdensity relative to the average value and is defined as 
\begin{eqnarray}
    \delta \equiv \frac{\rho}{\bar{\rho}} - 1 ~, \label{eq:dens_contr}
\end{eqnarray}
where $\rho$ is the local value of density and $\bar{\rho}$ is the average density across the entire spatial slice (with the average weighted by the proper volume). We note that given the strong inhomogeneities at late times, the average density is slicing dependent and cannot be tied to an FLRW background value. Nevertheless this provides a useful diagnostic to indicate where oscillons are forming, and track how the maximum density contrast changes over time. In particular, we identify that oscillons are formed when the maximum density contrast oscillates around some non-zero value as they decouple from the Hubble flow when $\delta \sim 1$. Note that what we refer to as the field density is split into the usual GR part for a minimally coupled scalar field, $\rho_\phi$, and a part arising from the modification to GR, $\rho_\mathrm{GB}$, where the latter is not a true energy density but the contribution of the curvature terms that appear in the equations of motion for the metric in the same way as a matter energy density. As a result this contribution does not obey any of the usual energy conditions. 
Explicitly:
\begin{subequations}\label{edgbcomp}
\begin{eqnarray}
\rho_{\mathrm{GB}}&=&\Omega M - 2M_{kl}\Omega^{kl}\,,
\end{eqnarray}
\end{subequations}
with
\begin{subequations}
\begin{eqnarray}\label{MNeq}
\hspace{-0.5cm}M_{ij}&=&R_{ij}+\tfrac{2}{9}\gamma_{ij}K^2+\tfrac{1}{3}KA_{ij}-A_{ik}A^k_j\,, \\
\hspace{-0.5cm}\Omega_i&=&-4f'(\phi)\big(D_i\Pi+A^j_{~i}D_j\phi+\tfrac{K}{3}D_i\phi \big) \nonumber\\
&&-4f''(\phi)\Pi\,D_i\phi\,,\\
\hspace{-0.5cm}\Omega_{ij}&=&4f'(\phi)\left(D_iD_j\phi+\Pi\,K_{ij}\right) \nonumber\\
&&+4f''(\phi)(D_i\phi) D_j\phi\,,
\end{eqnarray}
\end{subequations}
where $\Pi$ is the conjugate momentum of $\phi$, $\Omega=\gamma^{ij}\Omega_{ij}$, and $\Omega_i$ and $\Omega_{ij}$ come from the $3+1$ decomposition of the Weyl tensor.

The density contrast around a typical oscillon in a simulation is shown in Fig. \ref{fig:snapshot_rho_phi_v_GB}, along with the size of the EsGB modification to the density, relative to the average value on the slice. We see that it has regions with both positive and negative contributions.

\subsection*{Compactness}
The \textit{compactness} of an object is defined by the ratio of its gravitational mass to its radius, and gives a sense of how close to black hole formation the objects are,
\begin{eqnarray}
    \mathcal{C} \equiv \frac{M}{8 \pi R \mpl^2}, \label{eq:comp}
\end{eqnarray}
where $M$ and $R$ are the mass and radius of the objects, respectively. In the case of a black hole, $\mathcal{C}=1/2$. To compute the compactness of an oscillon, we first calculate its mass $M$ and proper volume $V$ as
\begin{equation}
    M = \int_\Omega d^3x \sqrt{\gamma}\rho\,, \qquad
    V = \int_\Omega d^3 x \sqrt{\gamma} \,,\label{eq:volume}
\end{equation}
where the spatial extent is defined as $\Omega = \{x^{i} \,\,|\,\, \rho(x^{i})/\rho_\text{max}(x^{i}) > 0.05\}$ and $\sqrt{\gamma} = \chi^{-3/2}$ is the volume factor related to the spatial metric. That is, we define the surface of the oscillons as the regions where the density $\rho$ is less than 5\% of the maximum density $\rho_\mathrm{max}$. Then, the radius of the oscillon is defined to be $R = \left(3 V/4 \pi \right)^{1/3}$. This definition is useful because the oscillons can be quite irregular in shape, and even at later times do not become spherically symmetric.

As with the density contrast, we can treat the mass as being the sum of a GR part and an EsGB part, $M_\mathrm{tot} = M_\mathrm{\phi} + M_\mathrm{GB}$. The evolution of $M_\mathrm{GB}$ and the compactness for different couplings are illustrated in the top panels of Fig. \ref{fig:oscillon_mass}.

\subsection*{Weak coupling condition}
\begin{figure}[t]
    \includegraphics[width =\linewidth]{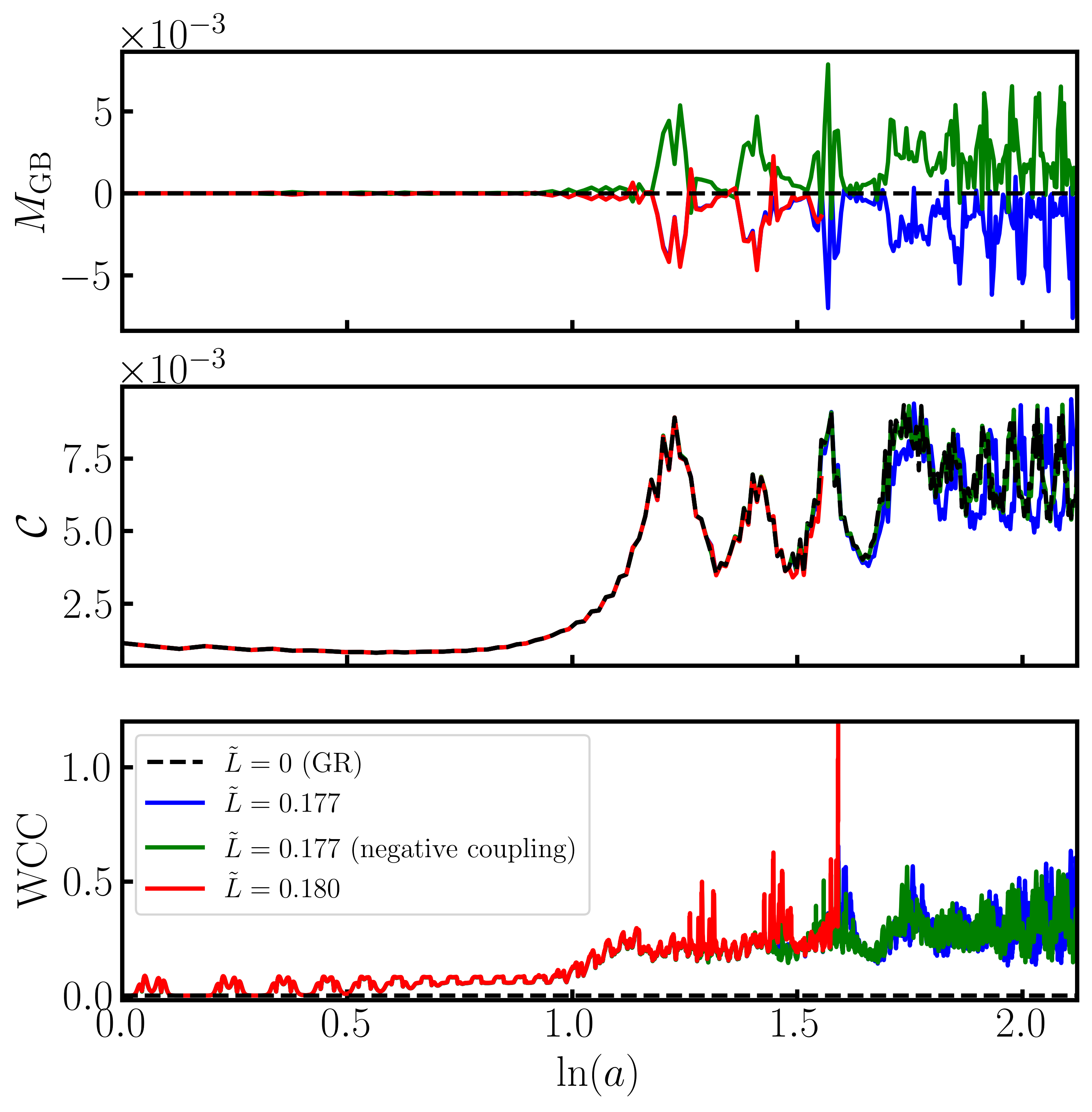}
    \caption{Evolution of the diagnostic quantities over time: The top panel shows the contribution to the effective oscillon mass from the Gauss-Bonnet corrections, which follows the sign of the coupling and remains at percent level even up to the critical cases studied at which the theory breaks down. The middle panel shows the oscillon compactness, which is not significantly changed by the modification. The bottom panel shows the weak coupling condition Eq. \eqref{eq:wcc}, which grows as the oscillon forms and smaller curvature scales become relevant.}
    \label{fig:oscillon_mass}
\end{figure}

The \textit{Weak Coupling Condition} (WCC) quantifies to what extent the spacetime is in the regime of validity of the EFT, where the corrections remain perturbatively small. It requires that the length scale associated with the scalar-Gauss-Bonnet coupling is small compared to the other relevant length scales in the system. The definition of the WCC given here is not covariant. One could use a covariant definition based on the relative size of the four-derivative terms compared to the two-derivative terms in the Lagrangian. However, this latter definition would not take into account possible differences between different directional derivatives, and hence, the definition here is more conservative. See Refs. \cite{Kovacs:2021lgk, AresteSalo:2022hua, AresteSalo:2023mmd, Doneva:2023oww, Brady:2025zxp} for more details.
The WCC requires that
\begin{eqnarray}
    \sqrt{|f'(\phi)|}/L \ll 1\,.\label{eq:wcc}
\end{eqnarray}
for all relevant length scales in the problem, such that $L$ is given by
\begin{equation}
L^{-1}
=
\max \left(
\begin{aligned}\label{eq:L_max}
& |R_{ij}|^{1/2} \\
& |\mathcal{R}_{\mathrm{GB}}^{2}|^{1/4} \\
& |\nabla_\mu \phi| \\
& |\nabla_\mu \nabla_\nu \phi|^{1/2} \\
& |V(\phi)|^{1/2}
\end{aligned}
\right) ~.
\end{equation}
Note that the dimensionless curvature scale $\tilde{L}$ that we use in the figures corresponds to the last case above ($|V(\phi)|^{1/2}$) evaluated at the start of the simulation (that is, it measures the coupling relative to the potential energy scale). As the simulation progresses, and the oscillon forms, other scales become more relevant, and the WCC can be violated dynamically. The evolution of the WCC for different couplings is illustrated in the bottom panel of Fig. \ref{fig:oscillon_mass}, and split out by component in Fig. \ref{fig:wccterms}.

\subsection*{Loss of hyperbolicity}
While the violation of the WCC indicates that the EFT has left its regime of validity defined in a strict sense, it may still be possible to evolve the EFT equations beyond the weakly coupled regime in some cases. This may be justified provided that the higher order derivatives neglected in the effective action remain small compared to the EsGB terms. The Cauchy evolution of EsGB theory can be extended (with a suitable choice of gauge) as long as the characteristic polynomial associated with the physical degrees of freedom is hyperbolic. Since EsGB theory is a scalar-tensor theory that propagates a scalar and two massless gravitational degrees of freedom (i.e. three physical degrees of freedom in total) and has second order equations, the characteristic polynomial associated with the physical degrees of freedom in this theory is of degree six. It was shown in \cite{Reall:2021voz} that this polynomial factorises into a quartic and a quadratic polynomial and therefore the propagation of gravitational and scalar waves in EsGB theory is determined by the corresponding quartic and quadratic cones. The ``fastest'' degree of freedom is always associated with the quartic cone (i.e. the quadratic cone always lies inside the quartic cone) and hence the causal structure and the hyperbolicity of the theory are primarily determined by the quartic cone. However, an efficient way to check the hyperbolicity of the quartic cone is currently lacking so in this paper we follow \cite{AresteSalo:2022hua,AresteSalo:2023mmd} and use the quadratic cone as an indicator for the breakdown of hyperbolicity. In particular, we will detect the breakdown of hyperbolicity using the criterion that the normalised determinant of the effective metric associated with the quadratic cone becomes negative. This effective metric is given by
\begin{align}
g_{\text{eff}}^{\mu\nu}=g^{\mu\nu}-4{\mathcal C}^{\mu\nu}\,,
\end{align}
where ${\mathcal C}^{\mu\nu}=\nabla^{\mu}\nabla^{\nu}f(\phi)$
and therefore the normalised determinant is calculated as
\begin{align}\label{eq:disc}
        \frac{\det(g^{\mu\nu}_{\text{eff}})}{\det(g^{\mu\nu})}=&~\tfrac{1}{(1+\Omega^{\perp\perp})^2}\det\left\{\tfrac{1}{\chi}\Big[(\gamma^{ij} - \Omega^{ij})(1 + \Omega^{\perp\perp}) \right. \nonumber\\& \left. -\tfrac{2}{\alpha}\Omega^{\perp(i} \beta^{j)} -\Omega^{\perp\perp} \tfrac{\beta^i \beta^j}{\alpha^2}+ \Omega^{\perp i} \Omega^{\perp j}\Big]\right\}\,,
    \end{align}
where $\Omega^{ij}=4\,\gamma^i_{\mu}\gamma^j_{\nu}{\mathcal C}^{\mu\nu}$, $\Omega^{\perp i}=-4\,n_{\mu}\gamma^i_{\nu}{\mathcal C}^{\mu\nu}$ and $\Omega^{\perp\perp}=4\,n_{\mu}n_{\nu}{\mathcal C}^{\mu\nu}$. When the value of the diagnostic in \eqref{eq:disc} becomes negative, the evolution equations are no longer strongly hyperbolic, and the theory breaks down as a predictive model. This can occur for high couplings within the dense oscillons that form during preheating, as can be seen in Fig. \ref{fig:deteff}.

\section{\label{sec:results}Results}

Here we briefly summarise the results in terms of the above diagnostics.

We observe many initial interacting proto-oscillons, which coalesce, stabilise, and decouple from the Hubble flow, with $O(10)$ oscillons forming in the simulation box (identified by tracking 95\% $\rho_{\mathrm{max}}$ isosurfaces). We focus our grid resolution entirely on studying a single isolated oscillon in detail, which at that point is unaffected by the others that have formed.
We tested a range of values and signs for the coupling. We located the critical strength of the coupling to be $\tilde{L} \sim 0.180$. Below this value, the formation of oscillons is mostly unaffected by the presence of the modifications. In Fig. \ref{fig:oscillon_mass} we see that the compactness for all cases is roughly consistent with the GR case, so that the modifications do not appear to increase (or decrease) the chance of the perturbations collapsing to form primordial black holes, and should not significantly affect the stochastic gravitational wave background they would generate.
The change in the effective mass of the oscillons is shown in the top panel of this figure, and it is of order $10^{-2}$ even in these critical cases, with a sign that follows the coupling $\lambda$.

\begin{figure}[t]
    \includegraphics[width =\linewidth]{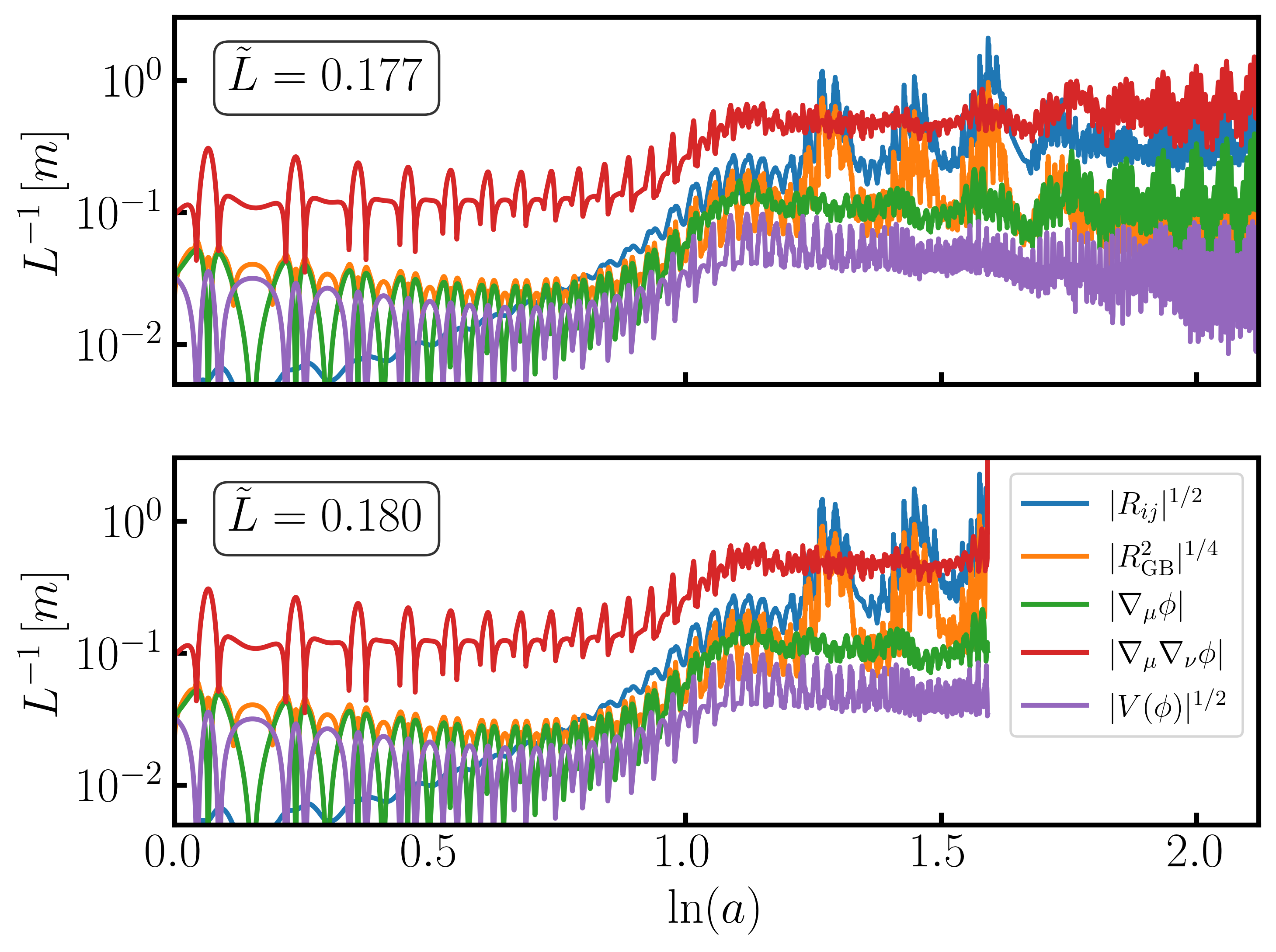}
    \caption{The different characteristic length scales of the weak coupling condition, Eq. \eqref{eq:wcc}. We see that in the critical case (lower panel) the growth of $R_{ij}$ and the second derivative of the scalar field drive the breakdown of the theory. In the sub-critical case (upper panel) the values stabilise and no breakdown is observed.}
    \label{fig:wccterms}
\end{figure}

\begin{figure}[t]
    \includegraphics[width =\linewidth]{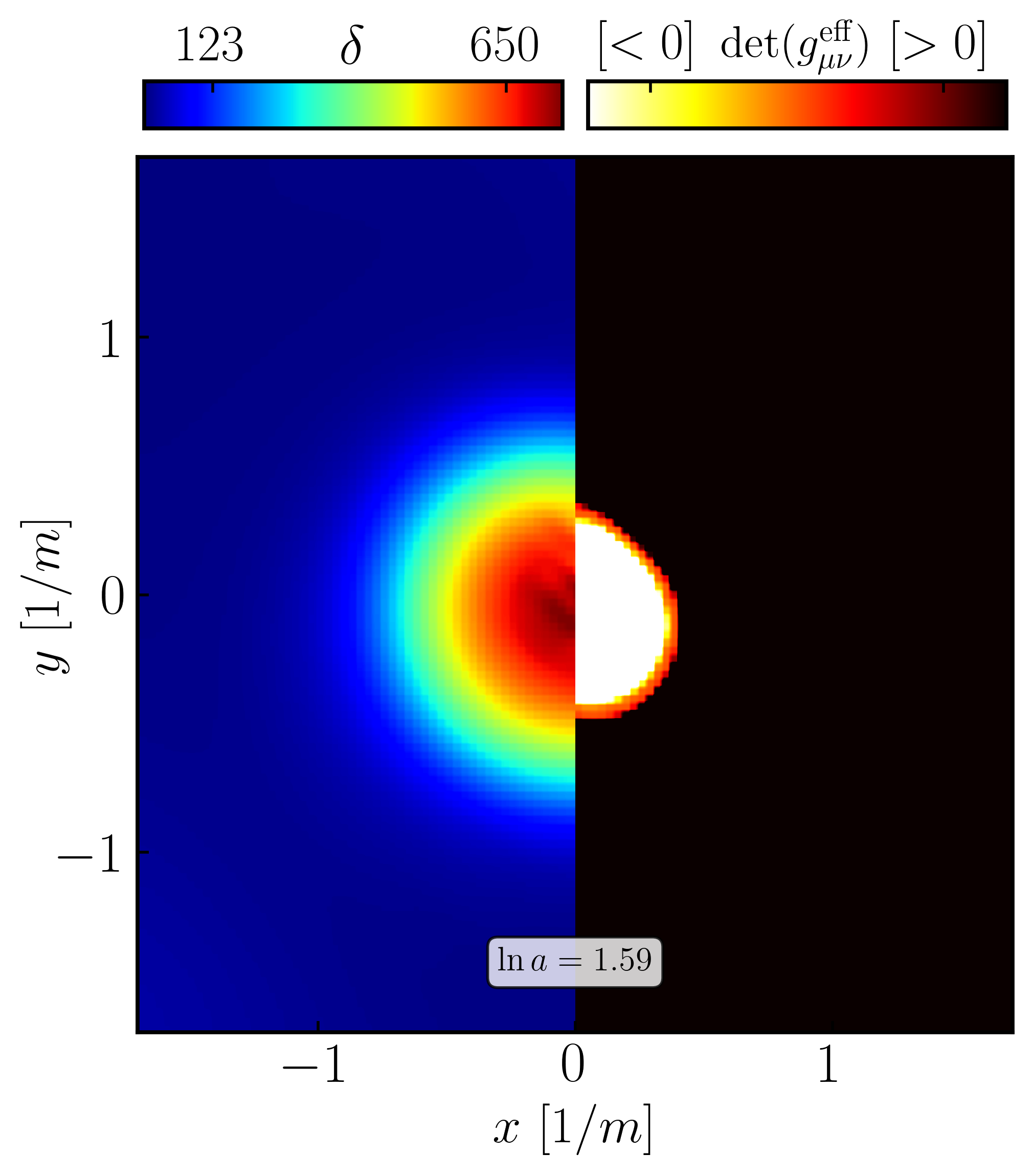}
    \caption{A 2D-slice from the simulation with positive $\lambda$ and $\tilde{L} \sim 0.180$ at the point of breakdown. The left half shows the density contrast, while the right depicts the determinant of the effective metric, which has become negative at the centre of the oscillon, indicating a loss of hyperbolicity.}
    \label{fig:deteff}
\end{figure}

In the case of a positive coupling, beyond the critical value $\tilde{L}\sim0.180$, we see that the weak coupling condition is violated, leading to a breakdown in strong hyperbolicity. In Fig. \ref{fig:wccterms} we analyse the different characteristic length scales in the problem compared to the coupling, and how they evolve over time. We see that the terms involved in the breakdown of the critical case are the curvature terms $R_{ij}$ and the second derivative of the scalar field, which are both maximum at the centre of the oscillon. Looking at the value of the normalised determinant of the effective metric (\ref{eq:disc}), we see that indeed this has become negative within the oscillon in the critical case, indicating a loss of hyperbolicity, see Fig. \ref{fig:deteff}. In the other cases the coupling remains large compared to the other length scales, but appear to stabilise. We found no evidence of a breakdown of the EFT when we ran several cases for longer.
We also tested negative couplings and found a very similar picture, only with a slightly higher value of the coupling strength associated with the breakdown of the EFT.

We can interpret these results in terms of the relevant length scales of preheating and the oscillons that form. The coupling strength $\tilde{L}$ is quoted relative to the reheating scale $H_\mathrm{reh}$, and whilst it is large, it remains below unity and is insufficient to break the EFT. Moreover, during the early stages of preheating, the universe expands, causing this scale to decrease and making an EFT breakdown even less likely.
However, when oscillons form, they decouple from the background expansion, introducing a new dynamical length scale set by their self-gravity. It is this scale that becomes the dominant one to compare against, and which ultimately breaks the EFT.

\section{Discussion}

The non-linear processes of preheating provide a well-motivated model for the formation of dense structures in the early universe, seeded only by primordial fluctuations at levels consistent with those on CMB scales. However, the mechanism is only efficient up to a point, and it has been shown in previous work that the simplest single-field models struggle to produce high compactnesses \cite{Aurrekoetxea:2023jwd}, and thus formation of primordial black holes through the mechanism seems unlikely in such scenarios. 

In this work we were motivated by the question of whether additional higher derivative couplings in the scalar-tensor system would change this conclusion. For modest couplings in which the system remains in the regime of validity of the effective field theory at all times, then broadly the answer is no. The properties of the oscillons that are formed (in particular their compactnesses) change little compared to the pure GR case, with some modifications sourced by the gradients of the field and localised at the oscillon walls.

However, in the case of larger couplings, we find that the dynamics of the formation of the oscillons and the resulting introduction of higher curvature scales due to their gravitational collapse can drive the evolution out of the range of validity of the effective field theory we are considering. At this point the Cauchy evolution breaks down, and we cannot follow to the end point of the evolution
\footnote{The key point here is really the invalidity of the EFT. The formalism used in \texttt{GRFolres} is only well-posed in the weak coupling regime and generically loses hyperbolicity upon entering the strong coupling regime \cite{AresteSalo:2022hua,AresteSalo:2023mmd}.While a better formulation might allow us to evolve further, usually such a loss of hyperbolicity in the theory is an artefact of the truncation of the true theory, and an indication that the EFT expansion has broken down, meaning neglected higher-order curvature terms would now significantly alter the dynamics.}.
This indicates that the derivative expansion also breaks down. Moreover, it tells us that preheating models that consider large couplings for EsGB models cannot be trusted to predict the end point of an evolution, and approximations that neglect non-linear and backreaction effects may break down in such cases. Note that the criteria ``large'' here needs to make reference to the curvature scales of the oscillons that form, rather than the background Hubble scale, since these are the terms that dominate during their formation.

In the case considered in this work, we set $g_2(\phi)=0$ and considered only the Einstein-scalar-Gauss-Bonnet contribution, but given the important contributions of the gradients of the field, we expect that these corrections should be large and therefore may also play an important role in the oscillon dynamics. We leave this question to future work.

In the absence of corrections at stellar mass black hole curvatures (of order $\sim$km), cosmological observations from the early universe may be the only way to probe higher curvature scales in gravitational EFTs. Our results suggest that if such effects do play a role on the length scales of inflation, their effect may also be significant in the subsequent non-linear dynamics of preheating. If corrections only enter at scales that are small relative to those of preheating, then GR provides a good description of the results, as expected.

\section{\label{sec:acknowledge}Acknowledgements}

We would like to thank Llibert Arest\'e-Sal\'o, Thomas Baumgarte, Sam Brady, Tom Giblin, Alan Guth, David Kaiser, Vincent Vennin and Serdar Yildiz for helpful conversations.
We thank the GRTL collaboration\footnote{\texttt{www.grtlcollaboration.org}} for their support and code development work. AW acknowledges the support of the Development and Promotion of Science and Technology Talents Project (DPST), the Institute for the Promotion of Teaching Science and Technology (IPST), Thailand.  JCA acknowledges funding from the Department of Physics at MIT through a CTP Postdoctoral Fellowship. KC is supported by an STFC Ernest Rutherford fellowship, project reference ST/V003240/1 and by the Simons Foundation International and the Simons Foundation through Simons Foundation grant SFI-MPS-BH-00012593-03. PF, AK and KC are supported by an STFC Research Grant ST/X000931/1 (Astronomy at Queen Mary 2023-2026). 
This work used the DiRAC@Durham facility managed by the Institute for Computational Cosmology on behalf of the STFC DiRAC HPC Facility (www.dirac.ac.uk). The equipment was funded by BEIS capital funding via STFC capital grants ST/P002293/1, ST/R002371/1 and ST/S002502/1, Durham University and STFC operations grant ST/R000832/1. DiRAC is part of the National e-Infrastructure. We also acknowledge the EuroHPC Joint Undertaking for awarding this project access to the EuroHPC supercomputer LUMI-C, hosted by CSC (Finland) and the LUMI consortium through a EuroHPC Regular Access call.
For the purpose of Open Access, the author has applied a CC BY public copyright licence to any Author Accepted Manuscript version arising from this submission.


\bibliography{main}

\appendix

\section{Numerical methods and convergence}

\begin{figure}[b]
    \includegraphics[width =\linewidth]{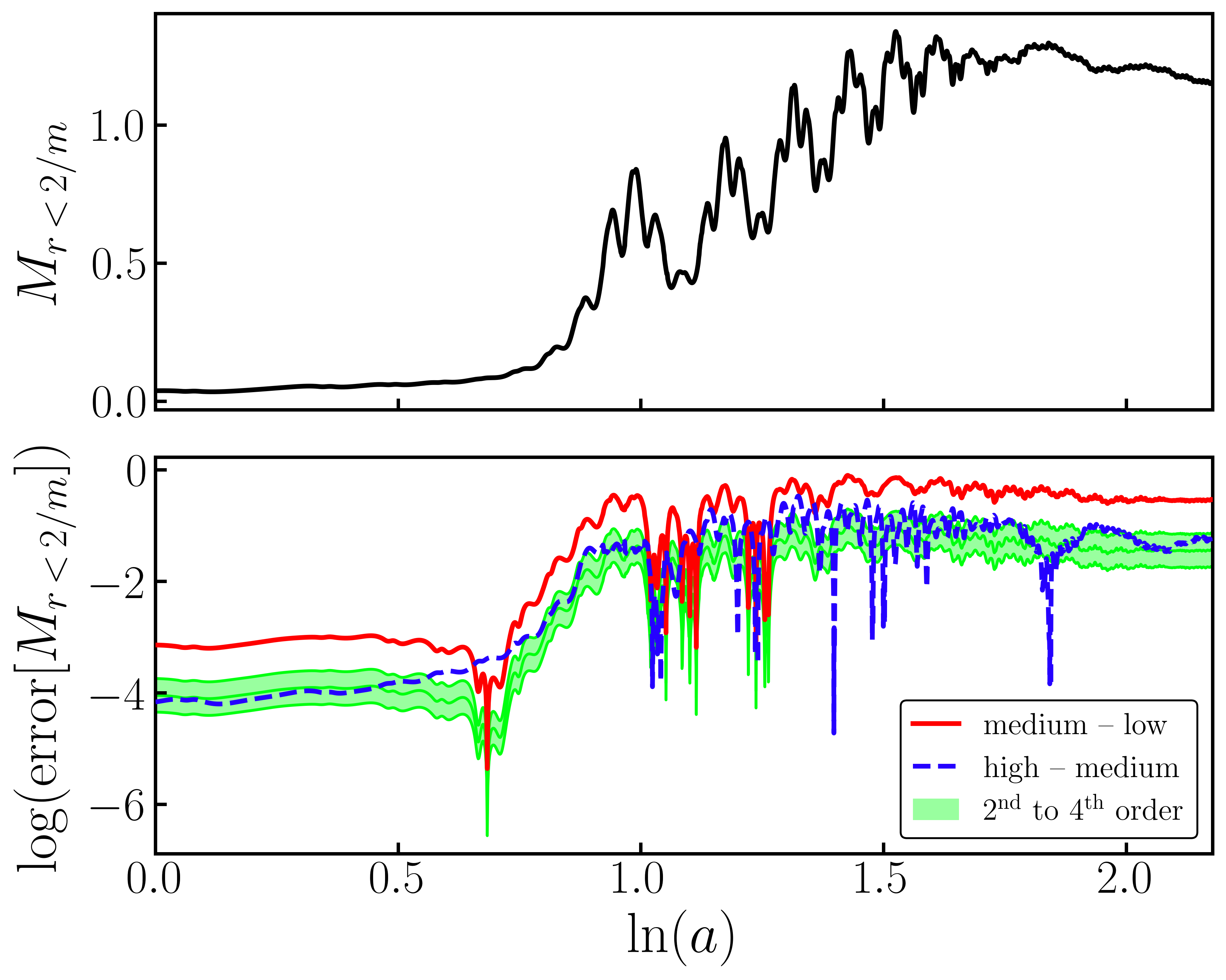}
    \caption{Convergence test with $N = \left\{64,\, 128,\, 256\right\}$ - low, medium and high resolution respectively. We test the evolution of a quantity related to the mass of the oscillon contained within a sphere of coordinate radius 2\,$m^{-1}$, which converges at 2nd to 4th order as expected. The top panel shows the measure used of the mass ($M_{r\,<\,2/m}$) over time. The bottom panel shows the errors and the expected convergence rate.}
    \label{fig:convergence}
\end{figure}

We evolve the coupled system of the EsGB equations using the standard CCZ4 formulation of \cite{Alic:2011gg} and the moving puncture gauge \cite{Bona:1994dr,Baker:2005vv,Campanelli:2005dd,vanMeter:2006vi} within the numerical-relativity codes \texttt{GRChombo} and \texttt{GRFolres} \cite{AresteSalo:2023hcp,Andrade:2021rbd,Radia:2021smk,Clough:2015sqa} 
, which use the method of lines, with an RK4 time integration and 4th order finite difference stencils for calculating spatial gradients. The initial data is generated using the code \texttt{GRTresna} \cite{Aurrekoetxea:2025kmm} using the CTTK method of \cite{Aurrekoetxea:2022mpw}.
We perform convergence tests to validate our simulations by repeating the runs at three different resolutions, with the number of grid points on the coarsest level $N = \left\{64,\, 128,\, 256\right\}$, with 5 levels of adaptive mesh refinement. 
We present the case where $\mu = 0.06\mpl$, $\langle\delta\phi\rangle^2 \approx 10^{-8}\mpl^2$ and $\tilde{L} = 0.162$ in Fig. \ref{fig:convergence}, which shows between 2nd to 4th order convergence as expected from the finite difference schemes used in the initial data solver (2nd order) and evolution code (4th order).
Here we define the error as the difference in the approximate oscillon mass $M$ compared between three different resolutions. For the purposes of this convergence test $M$ is evaluated as an integral of the energy density over the proper volume contained within a sphere of coordinate radius $r=2 m^{-1}$ centred on the oscillon coordinates $\{26.6406,\,22.7969,\,34.1406\}$. This quantity oscillates less than most of the diagnostic quantities that we use and so is better suited to checking the convergence. We expect the errors in our simulations to decrease at a predictable rate as we increase the resolution of the runs. We define the convergence factor,
\begin{equation}
    c(t) = \frac{|\phi_{\Delta_1} - \phi_{\Delta_2} |}{|\phi_{\Delta_2} - \phi_{\Delta_3}|} ~,
\end{equation}
where $\Delta_1,\,\Delta_2,\,\Delta_3$ are the spatial step sizes at the coarsest grid level for each resolution, from lowest resolution to highest. In the $\Delta \to 0$ limit,
\begin{equation}
    \lim_{\Delta \to 0}{c(t)} = \frac{\Delta_1^n - \Delta_2^n}{\Delta_2^n - \Delta_3^n}\,,
\end{equation}
where $n$ is the order of convergence.

We find good convergence in all cases studied up until the loss of hyperbolicity. The highest resolution triggers an earlier breakdown for $\tilde{L}=0.177$ (probably due to better resolution of the peaks of oscillations), so a lower coupling case $\tilde{L} = 0.162$ is displayed in Fig. \ref{fig:convergence} to show a fully comparable evolution for the time period displayed in the plots of the main text.

\end{document}